\documentclass[a4paper,11pt]{article}
\usepackage{jinstpub}
\usepackage{lineno}
\usepackage[outdir=./]{epstopdf}
\usepackage{multicol}
\usepackage{multirow}
\usepackage{longtable}
\usepackage{array}
\usepackage{rotating}
\usepackage{arydshln}

\usepackage{ptdr-definitions}
\usepackage{custom-definitions}

\usepackage{enumerate}

\newcommand{\orc}{\includegraphics[height=\fontcharht\font`A]{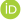}}
\newcommand{\orcid}[1]{\href{https://orcid.org/#1}{\orc}}

\title{Conceptual study of a two-layer silicon pixel detector to tag the passage of muons from cosmic sources through quantum processors}

\author{Ula{\c s}can Sar{\i}ca\orcid{0000-0002-1557-4424}}
\affiliation{University of California, Santa Barbara\\Santa Barbara, CA, USA}

\emailAdd{ulascan.sarica@cern.ch}

\abstract{
Recent studies in quantum computing have shown that quantum error correction with large numbers of physical qubits are limited by ionizing radiation from high-energy particles. Depending on the physical setup of the quantum processor, the contribution of muons from cosmic sources can constitute a significant fraction of these interactions. As most of these muons are difficult to stop, we perform a conceptual study of a two-layer silicon pixel detector to tag their hits on a solid-state quantum processor instead. With a typical dilution refrigerator geometry model, we find that efficiencies greater than 50\% are most likely to be achieved if at least one of the layers is operated at the deep-cryogenic ($<1\K$) flanges of the refrigerator. Following this finding, we further propose a novel research program that could allow the development of silicon pixel detectors that are fast enough to provide input to quantum error correction algorithms, can operate at deep-cryogenic temperatures, and have very low power consumption.
}


\begin{document}
\maketitle
\flushbottom

\section{Introduction}
\label{sec:Intro}

Quantum computers have recently been shown to perform calculations that would be practically impossible with classical supercomputers~\cite{Arute:2019zxq,Zhong:2020}. These computers greatly benefit from quantum error correction (QEC) algorithms that couple the states of an arbitrary number of $k$ logical qubits over $n>k$ physical qubits: Physical qubits may have error rates at the level of $10^{-3}$ individually, but one can exploit their small and mostly decorrelated nature to determine the original logical states more precisely. Increasing the number of physical qubits per logical qubit can therefore reduce the logical error rates exponentially~\cite{Chen:2021,Acharya:2023}.

In repetition code studies using solid-state devices, however, it has been shown that this exponential reduction flattens beyond a code distance of 20, down to a level of $10^{-6}$ per correction cycle when using cyclical corrections~\cite{Acharya:2023}. Ionizing radiation from high-energy particles, which can come from cosmic rays, or natural radiation from the environment and the materials used in electronics, are understood to be one of the contributors to this flattening~\cite{Grunhaupt:2018,Wilen:2020lgg,Vepsalainen:2020trd,McEwen:2021wdg,Cardani:2020vvp,Cardani:2022blq}.

Briefly, as a high-energy particle passes through the quantum processor, it can induce quasiparticles in the carrier chip. Without any mitigation efforts, these quasiparticles could spread over a large surface area and disrupt coherence across the qubits in the area neighboring the particle's hit position.
Detailed measurements on Google's Sycamore processor~\cite{Arute:2019zxq}, which features a $20\mm\times26\mm$ carrier chip and a $(10\mm)^2$ qubit chip, document this process to occur randomly at a rate of $0.1\Hertz$ on average. Currently, once a high-energy particle hits this processor, the occurrence of errors quickly spreads across multiple qubits within as little as $10\mus$ (corresponding to an increase from a reported baseline of 4 simultaneous qubit errors to 10), and the error rate decays exponentially with a decay constant of $25\ms$~\cite{McEwen:2021wdg}.

Current rates of correlated quantum errors due to high-energy particles are more likely to be dominated by natural radiation than cosmic sources, but their proportions depend on the physical location of the quantum system, \ie, the amount of rock and concrete above the system, and the natural radiation properties of the electronics and the surrounding environment. Measurements of radioactivity in the laboratory environment and further simulation studies of the Fermilab-SQMS/INFN-LNGS Round Robin chip show that about 64\% of particle interactions would be due to $\gamma$ radiation in an above-the-ground facility, whereas about $36\%$ would be due to muons from cosmic rays, followed by a trace amount of contribution from neutron radiation. In comparison, rates at the $1.4\km$-deep LNGS facility are calculated to be almost exclusively due to $\gamma$ radiation as the thick rock layer acts as a barrier to muons~\cite{Cardani:2022blq}. Particle interaction rates similar to the ground-level estimates of this study were also reported in Ref.~\cite{Wilen:2020lgg}.

Strategies considered so far to mitigate the effect of high-energy particles include improving the design of chips with phonon and quasiparticle traps or membrane structures~\cite{Wang:2014,Nsanzineza:2014,Daddabbo:2014,Henriques:2019,Karatsu:2019,Martinis:2021}, adding shielding from natural radiation~\cite{Cardani:2022blq}, distributing QEC across separate chips~\cite{Xu:2022}, and using co-located transition-edge sensors (TES) to detect energy transmission into the qubits from ionizing radiation~\cite{Orrell:2021}:
\begin{itemize}
\item The introduction of phonon and quasiparticle traps, or membrane structures, could help reduce the sensitivity of the quantum processor to sudden bursts of energy from high-energy particle interactions. These implementations could also help limit the spread of the energy burst into several qubits.
\item Additional shielding surrounding the quantum processor could protect it from natural sources of $\gamma$ rays and neutrons. However, muons from cosmic sources are typically more energetic, and as also noted in Ref.~\cite{Cardani:2022blq}, the same shielding equipment would be unable to stop most muons.
\item A distributed QEC scheme relies on the availability of a network of processors to improve the fault-tolerance over certain types of errors. Implementing a network architecture might also require more complex qubit control schemes and electronics, so more experimental investigation needs to be performed to assess its feasiblity using the current technology.
\item Transition-edge sensors operate at deep-cryogenic temperatures to detect very small changes in temperature induced by the passing particles as the sensors switch from being superconductive to normal conductive. If these devices can be placed adjacent to the quantum processor, they could help reject calculations that are marked by these detectors as erroneous, or mark the affected qubits directly in the case sensors are uniquely associated to each qubit. However, as Ref.~\cite{Orrell:2021} also notes, the rising-edge of the reaction time of TESs are typically around $5\mus$, which is comparable to the rising edge of error occurrence in the Sycamore processor and about five times slower than the QEC round time~\cite{McEwen:2021,McEwen:2021wdg}. Therefore, marking individual qubits for errors using TESs, even if through the identification of a rising edge in the electronic signal, requires reductions in TES reaction times.
\end{itemize}
Reducing error rates, in the end, is a continuously developing process, and multiple approaches will ultimately need to be combined together.

In this paper, we propose to supplement these techniques by commissioning a two-layer silicon pixel detector setup to detect the muons passing through the qubit chip. We choose silicon as sensor medium as it will have high detection efficiency for momentum ranges relevant for muons from cosmic sources, and typical silicon detectors used in particle physics have time resolutions at least an order of magnitude less than the QEC round time. With a fine enough pixellation and suitable distances between the two detector layers, and between each layer and the qubit chip, it is possible to make a linear muon track prediction and identify the particle hit on each qubit with sufficient accuracy and efficiency. Once the hit position on the qubit chip is identified, the corresponding qubit, or the qubits in the affected area, could already marked as erroneous, and the error correction algorithm could then bypass the marked set of qubits until they recover.

The paper is organized as follows: In Section~\ref{sec:MuonEffStudy}, we describe a simple simulation study for the passage of muons through the dilution refrigerator, and outline our observations and conclusions from different possible pixel sizes and detector layer configurations. The numerical results supporting this study are presented further in Appendix~\ref{app:sec:MuonEffStudyDetails}. In Section~\ref{sec:RDChallenges}, we overview some of the challenges we anticipate in building such a detector for deep-cryogenic operation and offer possible solutions to investigate, and, finally, in Section~\ref{sec:Discussion}, we suggest possible steps for further research and discuss its synergistic benefits.

\section{Muon detection efficiencies with a two-layer pixel detector setup}
\label{sec:MuonEffStudy}

Before a muon from cosmic sources reaches the qubit chip, it will have to go through the material of the dilution refrigerator. As it goes through each component, in most cases the copper or steel disks held at different temperatures, or the support structures holding these flanges, it is going to be deflected at small angles predominantly due to Coulomb scattering from the nuclei in the material. Using multiple scattering theory over thin media~\cite{PDG:2022pth}, one can calculate that muons within a momentum range of $0.5$--$10\GeV$ passing through $8\mm$ of copper (55.7\% of radiation length) at normal incidence are expected to scatter according to an approximately Gaussian distribution with a width of $1.2$--$0.057\degrees$. Consecutive flanges in a dilution refrigerator are roughly $200\mm$ apart in the geometry model we study, so deflections on the surface of each succeeding flange are expected to be about $4$--$0.2\mm$.

These deflection distances on the flange surfaces are comparable to the current qubit dimensions of $1\mm$ in the Sycamore processor~\cite{Arute:2019zxq,McEwen:2021wdg}, so including rudimentary multiple scattering effects in a simulation study is important to understand the accuracy of a qubit hit position prediction from a linear trajectory between two detector layers, especially if there is at least one flange in between them. While more detailed studies using the \GEANTfour package~\cite{Agostinelli2003250} could be performed, as done for example in Ref.~\cite{Cardani:2022blq}, to include further effects from natural radiation sources and secondary particle production from muon scattering, a simple numerical simulation that includes multiple scattering effects and particle propagation over a simplified dilution refrigerator geometry should suffice to assess the efficiency and accuracy of the different choices for pixel sizes and detector layer positions considered in this conceptual study.

\begin{figure*}[htb]
\centering
\includegraphics[width=0.6\textwidth]{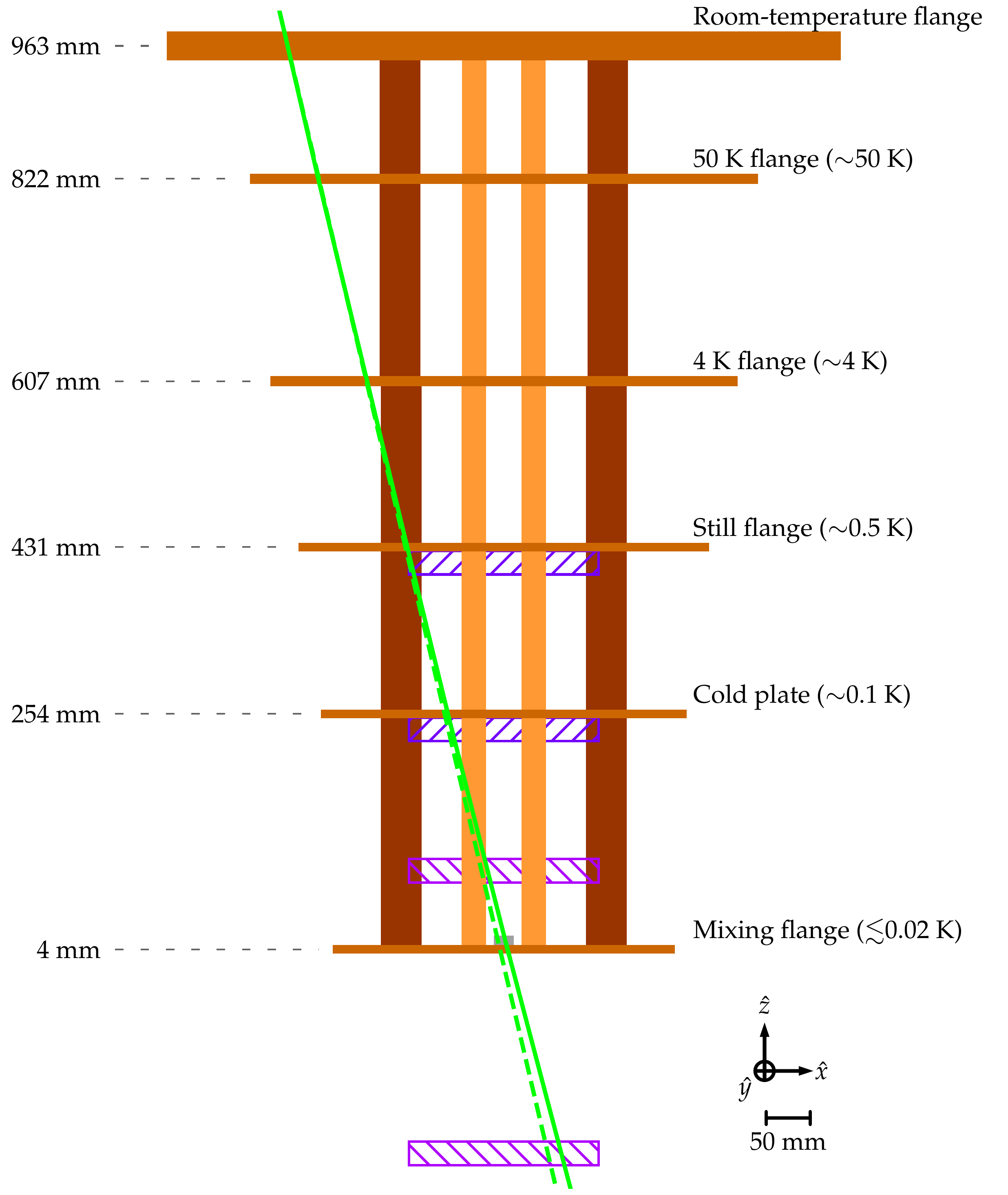}
\ccaption
{Two-dimensional sketch of the experimental setup}
{
The $zx$ projection of the setup considered for the study of muon tagging efficiencies is shown with the $y$ axis going into the page.
The dilution refrigerator components are depicted to scale in brown color with lighter and darker shades for support structures closer to and farther away from the page, respectively.
The naming convention of each disk is written explicitly on the right-hand side, along with the approximate operating temperatures in parentheses except for the room-temperature flange. The vertical positions of the center of disks, starting from the bottom of the mixing flange, are also displayed on the left, and the matching gray, dashed lines are shown to guide the eye.
The quantum processor is represented with the smallest gray rectangular box surrounding the qubit and carrier chips, and two possible positions of the top (bottom) detector layers are shown in blue (purple) open boxes hatched with forward (backward) slashes. The widths of the quantum processor and the detector layers are inflated by a factor of 100 for easier view. The broken, green, solid line illustrates the path of a muon hitting the qubit chip after multiple scattering from the dilution refrigerator components, and the straight, green, dashed line illustrates the path this muon would take without multiple scattering.
}
\label{fig:geometry}
\end{figure*}

The dilution refrigerator is modeled in a simplified manner as six (solid) disks and eight support structures (solid rods) based on the technical specifications of a commercial supplier. The geometrical and material specifications of these components are listed in Table~\ref{table:dilution-refrigerator}, and the sketch of the setup is illustrated in Fig.~\ref{fig:geometry}. For simplicity, we assume that the disks, and the detector and quantum processor elements are centered around the axis common to all disks, and that a maximum area of $(200\mm)^2$ is available to place the detector layers. We note that different dilution refrigerator models might feature smaller or larger surface areas for experimentation or flange-to-flange distances by roughly a factor of two, so the available surface area we chose represents a mid-size dilution refrigerator model. Incidentally, these dimensions would also allow the silicon sensors to be manufactured using a single run, thanks to the availability of silicon wafers $300\mm$ in diameter.

\begin{table}[htbp]
\centering
\ccaption
{Geometry and material details of the dilution refrigerator model}
{
The specifications of the disks available to place detector layers and the quantum processor, and the support structures, modeled as solid rods, are combined in this table. The symbols $D$, $L$, and $z$ stand for the diameter, thickness (length), and the vertical position of the geometric center of the disks (rods), respectively. For support structures, the coordinate $r$ is the radial distance of the rod from the central symmetry axis of the disks, and the listed azimuthal angles, $\phi$, correspond to pairs of support structures of each type. The acronym ``RT'' in the naming of the uppermost disk stands for ``room temperature'', and all disks except the room temperature flange are assumed to be made of copper. The material for the room temperature flange, steel, is approximated with iron as the alloy consists of about 73\% iron~\cite{Carter:1994} and the rest mostly of other metals with similar \dEdxtxt properties~\cite{PDG:2022pth}.
}
\renewcommand{\arraystretch}{1.25}
\begin{tabular}{ c c c c c }
\hline
\multicolumn{5}{c}{Disks} \\
\hdashline
Disk name & $D$ ($\mmns$) & $L$ ($\mmns$) & $z$ ($\mmns$) & Material \\
\hline
RT flange & 710 & 30 & 963 & Fe \\
50\K flange & 535 & 10 & 822 & Cu \\
4\K flange & 492 & 10 & 607 & Cu \\
Still flange & 432 & 8 & 431 & Cu \\
Cold plate & 385 & 8 & 254 & Cu \\
Mixing flange & 360 & 8 & 4 & Cu \\
\hline
\multicolumn{5}{c}{Support structures} \\
\hdashline
$D$ ($\mmns$) & $L$ ($\mmns$) & $r$/$z$ ($\mmns$) & $\phi$ ($\degrees$) & Material \\
\hline
42 & 336 & 155/780 & 45 and 135  & Cu \\
25 & 336 & 134/780 & -76.5 and -103.5  & Cu \\
42 & 594 & 153/305 & 45 and 135  & Cu \\
25 & 594 & 135/305 & -76.5 and -103.5  & Cu \\
\hline
\end{tabular}
\label{table:dilution-refrigerator}
\end{table}

The modeling of the quantum processor features a geometry similar to that of the Google Sycamore processor: The carrier chip is represented with a $20\mm\times26\mm\times100\mum$ box, and the qubit chip is represented with a $12\mm\times12\mm\times100\nm$ box. For simplicity, we assume both components are made of silicon, but the exact details of the material composition should not have significant impact on multiple scattering due to the thinness of these layers. The carrier chip is assumed to lay directly on top of the mixing flange, and the qubit chip is assumed to lay directly on top of the carrier chip.

The vertical placement of the top detector layer determines the angular acceptance, and the only disks that can provide support for this layer with $\gtrsim50\%$ efficiency are the cold plate and the still flange. To avoid extra scattering before determining the muon trajectory, we always assume that the top detector layer can be attached directly underneath these disks.
A sufficiently large distance is needed between the top and bottom detector layers in order to be able to make an accurate trajectory prediction, but accuracy also improves with closer proximity to the qubit chip. Therefore, as also depicted in Fig.~\ref{fig:geometry}, we test for cases in which the bottom layer is placed either below but away from the cold plate, or over the ground and below the mixing flange. The former choice could achieve slightly better trajectory accuracy with the least amount of multiple scattering between the two layers while the latter allows the operation of the bottom layer at close to room temperature.

Each detector layer is modeled as a two-dimensional grid of square silicon pixels with $200\mum$ thickness. Neither their material composition nor their exact thickness are expected to contribute to multiple scattering significantly, so different assumptions that result in the same two-dimensional pixelation would lead to the same conclusions as far as this simulation study goes. The width of the pixels in the top layer is varied from $150\mum$ to $1000\mum$ in the tests performed, but the number of pixels are always adjusted to fill a maximum area of $195\mm\times200\mm$, leaving a space of $2.5\mm$ for readout-specific elements on either end along the $x$-axis, similar to an example readout chip design for the CMS pixel detector~\cite{CMS:1997tlf} at the CERN Large Hadron Collider (LHC). The full pixel area is assumed to be active.

When the bottom detector layer is between the top layer and the qubit chip, we keep the number of pixels the same as that of the top layer, and consider two possible scenarios: The sizes of each pixel could be kept the same as well, or they could be adjusted to match the projection of the top layer at the vertical position of the bottom layer. The same two scenarios are considered when the bottom layer is below the mixing flange as long as the layer is close enough that the maximum area of $195\mm\times200\mm$ still covers at least the same angular opening the top layer does. Once the bottom detector is so far below that this condition fails to hold, the former scenario requires an increase in the number of pixels at each dimension to match the angular opening; such an increase could be accommodated since we would not expect the bottom layers to be subject to the operational constraints of the dilution refrigerator for such placements.

The ultimate figure of merit in this study is the efficiency to detect muons. We define this quantity as the ratio of the number of muons that pass through both layers of the detector setup with valid hit predictions over the qubit chip surface to the total number of muons that hit the qubit chip. The linear approximations of the muon trajectories are built between the geometrical centers of the pixels through which the muons pass, and the predictions for the qubit chip hit position are determined from the projections of these trajectories on this chip.

Since reducing the spread of quasiparticles and phonons into multiple qubits is an active area of development, the size of this spread drives the ultimate accuracy requirement for hit position resolution in detector design. The exact placement of qubits, however, could be different depending on the details of the qubit chip. Therefore, in order to gain insight into muon detection efficiencies for different hit position accuracies, we simply assume that the qubit chip surface area is divided into a densely-packed $12\times12$ grid of qubits, taking the surface area of each qubit to be $(1\mm)^2$ based on the Sycamore processor. This assumption allows us to assess our numerical results, presented in greater detail in Appendix~\ref{app:sec:MuonEffStudyDetails}, for position accuracies corresponding to $1\times1$ (at the same qubit hit), $3\times3$ (within nearest neighbors of the qubit hit), or $5\times5$ (within next-to-nearest neighbors of the qubit hit) qubit cells.

For configurations with the top detector layer underneath the cold plate, muon detection efficiency is found to reach up to 77\%. Smaller pixel sizes naturally yield better overall detection efficiency, but adapting the pixel size in the bottom layer recovers part of the loss in single-qubit hit efficiencies, by as much as 20\% when the largest pixel sizes are used. In fact, when the bottom layer is placed closely above the qubit chip ($150\mm$ below the cold plate or, equivalently, $92\mm$ above the qubit chip), adapting bottom layer pixel sizes can yield muon detection efficiencies as high as 64\% for a reasonable top layer pixel area of $(500\mum)^2$.
Increasing the layer-to-layer distance (arm length) reduces single-qubit hit efficiencies, but efficiencies over $3\times3$ or larger qubit cells remain at maximal value.

For the same top detector layer positioning, placing the bottom detector layer below the mixing flange yields single-qubit hit efficiencies above 50\% only when the bottom layer is within about $100\mm$ of the bottom surface of the mixing flange. Considering the need to budget additional space for the vacuum enclosure and radiation shielding in the dilution refrigerator apparatus, we expect single-qubit hit efficiencies to be capped around this value.
We also observe the effect of multiple scattering from the mixing flange in two different ways: For the same top detector layer positions and pixel sizes, the bottom detector layer either has to be closer to the qubit chip or needs smaller adapted pixel sizes to maintain the single-qubit hit efficiency values that match the scenarios in which the bottom layer is placed below the cold plate and above the qubit chip instead.
Furthermore, when the arm length is long enough, multiple scattering from the mixing flange can further cause efficiencies within $3\times3$ and $5\times5$ qubit cells to fall below the maximal value of 77\% regardless of the pixel size.
We nevertheless note that efficiencies with an accuracy within a $3\times3$ qubit grid cell all remain above 74\% in the cases we tested. For this reason, the provided configurations still remain to be good potential targets for intermediate stages in technological development, especially when qubit hit resolution can be broader than the width of a single qubit.

The narrower angular acceptance in configurations with the top detector layer underneath the still flange reduces the maximum achievable efficiency to 52\%. Part of the reduction in efficiencies also comes from the fact that the cold plate acts as a source of multiple scattering in the cases where the arm length is long enough. While single-qubit hit detection efficiencies are significantly lower in all examined configurations, conclusions on the overall behavior of efficiencies remain qualitatively similar. The placement of the top detector layer beneath the still flange could therefore also be considered as an additional detector development stage.

\section{Current technologies and anticipated challenges}
\label{sec:RDChallenges}

Typical pixel detectors in particle physics experiments currently consume around 40~\cite{ATLAS:1998yql} to 60~\cite{CMS:1997tlf} $\muWattsns$ per pixel, and are limited mainly by readout electronics, with the majority of the power consumption due to amplifiers. Naively taking these values as a baseline yields a power consumption estimate of no less than 1.6\Watts per layer if we use one of the $(1000\mum)^2$-pixel configurations from Section~\ref{sec:MuonEffStudy}.
The cooling power of a dilution refrigerator, on the other hand, is around 1\Watts at 4\K and reduces exponentially at lower temperatures. The still flange is typically held at around $0.5\K$, and the cold plate is meant to act as a stepping stage for a reduction of the temperature down to $\sim0.02\K$~\cite{Arute:2019zxq} at the mixing flange.

Based on the low power requirements, the pixel detector approaches utilized in particle physics experiments are not yet sufficient for a full detector layer to be operable at the still flange or the cold plate, but there are techniques that are already in continuous development that could be helpful starting points for a low-power, deep-cryogenic detector. A few examples are as follows:
\begin{itemize}
\item At the LHC, proton-proton collisions occur every $25\ns$, and pixel detectors used in the LHC experiments need to be able to detect a large number of charged particle tracks with very high efficiency values within this time interval. Depending on the physical location of the quantum processor, the frequency of detecting hits on a single $(200\mm)^2$-pixel detector layer, on the other hand, would be between 8 and 40\Hertz~\cite{McEwen:2021wdg,PDG:2022pth,Cardani:2022blq}, orders of magnitude smaller than rates at the LHC.
With a time resolution of $\lesssim100\ns$ in the pixels layers, \ie, at least an order of magnitude smaller than the QEC round time, the fraction of concurrent $n>1$ muon hits at each detector layer would then be $<2\times10^{-6}$.

An intermediate step in readout development could therefore feature extracting the row and column information instead of individual pixels, without worrying about ghost hits that would arise from concurrent hits at a layer. This approach is already tested in the AstroPix v2 HV-CMOS sensors~\cite{Brewer:2021mbe,Steinhebel:2023vep}, which are pixel sensors with larger dimensions adapted from the ATLASPix developments~\cite{Peric:2007zz,Peric:2021bcu}. Readouts corresponding to the OR of the pixels on each column and row could still give the two-dimensional detector hit position, and further requirements to match the time of the row and column hits, and time-over-threshold (ToT) measurements could be employed to improve hit identification accuracy.

A related technology choice could be to directly utilize double-sided strip sensors, or single-sided strips that are placed close to each other orthogonally. The use of strip detectors would reduce the channel count and power consumption in readout drastically if they can be shown to perform well at temperatures below $1\K$. Capacitance in these detectors would need to be kept low; otherwise, increased capacitance could lead to worse signal-to-noise ratio, and there could be more combinatorial confusion in the presence of multiple hits, either due to noise, or other sources such as natural radiation or secondary particle interactions.
\item The power load from readout could also be reduced by connecting multiple pixels to a readout chip via AC coupling, translating their address into different pulse heights using adjustable gains, and encoding this information into a ToT signature. Alternatively, the row and column position of each pixel could be encoded through voltage or current drops in a strip detector-like readout. Both of these options have been tested to reduce effective pixel sizes in the ATLAS Inner Tracker~\cite{Miucci:2014jxa,Peric:2014faa,Hirono:2016zck,Garcia-Sciveres:2017ymt,ATLAS:2017svb}.
\item Another technique utilized in cosmic microwave background surveys~\cite{Barron:2017kuo,CMB:SPIE2022,CMB-S4:2022jnk} is staged amplification and readout. While these detectors use TES bolometers, and, therefore, the sensor and readout systems do not directly apply to this study, the basic idea of amplifying signals in multiple steps, starting with lower gains at the lowest temperatures, could be investigated to adapt for pixel detectors.
\end{itemize}

We note that the first two of the above pixel readout techniques rely on the use of HV-CMOS pixel sensors. These sensors have drift-driven charge collection, and are designed to have thinner depletion regions with bias voltage values within $<100\Volts$, even as low as $20\Volts$ based on some of the tests of ATLAS prototypes~\cite{Hirono:2016zck}. Considering that radiation damage is not expected to be a problem to detect muons from cosmic sources around the quantum processor, the bias voltage would not need to be increased beyond the requirements for a non-irradiated sensor to recover charge collection efficiencies. These pixel sensors also allow in-pixel front-end readout, which lowers power dissipation. The power dissipation ultimately desired in the development of these sensors are as low as $1\mWattsocmsq$~\cite{Brewer:2021mbe}.

The deep-cryogenic operability of the sensors is an equally crucial design aspect in this presented application. Solid-state detectors have been operated at sub-Kelvin temperatures with different purposes for decades. The CDMS Experiment, for example, has utilized undoped silicon and germanium sensors of about $1\cm$ thickness at temperatures $\lesssim0.05\K$~\cite{CDMS:2002moo,CDMS:2005jsf}. At these low temperatures, there are practically no mobile carriers in the crystals, so the sensors do not need to be operated in reverse-biased mode to create a depletion region. Indeed, in CDMS, charge collection has been achieved with small electric fields of a few $\Voltsocmns$ across the crystals and a low-noise readout system,
albeit with hourly cycles to recover ionized impurity sites from charge trapping, which might not be ideal for long quantum computations, and
typical readout time constants larger~\cite{Akerib:2008zz} than QEC round times in the adopted experimental design.

In the presence of doping profiles, carrier freezeout effects that occur at temperatures $<100\K$ become an important factor to control.
Recent tests on non-irradiated $p^+$--$n$--$n^+$ silicon sensors for possible high-luminosity LHC beam loss monitor designs~\cite{Eremin:2018vwc} show only small differences in current pulses at temperatures between $100\K$ and $7.6\K$ for 4.5 and $10\kOhmcm$ silicon sensors with a bias voltage of $100\Volts$ (above depletion voltage). The pulse shapes at these temperatures are narrower as a function of time than those at room temperature, but the collected total charge is only less than that at room temperature by $\sim3\%$. In contrast, significant degradations in current pulse shapes and total accumulated charge were found for sensors with a resistivity of $0.5\kOhmcm$, albeit with the caveat that the tested bias voltage in those cases depleted the sensors only partially.
While these observations may not necessarily hold exactly for HV-CMOS sensors from the discussion above, which feature $n$-wells in a $p$-substrate for the depletion layer, the findings might indicate that dopant levels in these types of sensors may also need to be fine-tuned in order to achieve better charge collection performance at cryogenic temperatures.

In the case of CMOS sensors, another related challenge for deep-cryogenic operation is the usability of electronics in these devices at sub-Kelvin temperatures. The detailed properties of the electronics depend on target applications, and circuits designed to operate above 100\K may not perform well below 1\K. Numerous studies in the literature have been focusing on characterizing CMOS electronics designed for operation at around $4\K$, and CMOS control and readout devices have been built to operate at these temperatures in quantum computing applications~\cite{Coskun:2014,Hornibrook:2014aba,Shin:2014,Sebastiano:2017,Balestra:2017,Beckers:2017,Bohuslavskyi:2017,Incandela:2018,Park:2018,Bardin:2019,Charbon:2019,Bonen:2019,Patra:2020,Sebastiano:2020,Xue:2021,Ghibaudo:2021,Hart:2022xex,Xue:2023}, or recently even as low as $\sim0.05\K$ for use with quantum dots~\cite{Ruffino:2022}. It would be instructive for the future experimental studies of the detector proposed in this paper to test simple CMOS circuits designed for operation at 4\K at the lower temperatures of the cold plate and the still flange. If their performance is unsatisfactory, it may be necessary to connect the sensors to readout elements located at higher temperatures, thereby applying staged amplification and readout techniques as mentioned above. In such a case, dedicated interconnects with very low thermal conductivity would need to be utilized.

The design of a detector also requires support structures. Here, we particularly consider the module base and external supports, and leave other technical details of the module assembly to a future study. The module base material should be able to transfer the dissipated power away from the silicon pixels, withstand any thermomechanical stress and shear, and should have a thermal expansion coefficient ($\alpha$) close enough to that of silicon to avoid inducing shear comparable to the pixel size, \ie, for pure silicon, the values of $\alpha$ range from $1.33\times10^{-9}\Kinv$ at 14\K to $-47.40\times10^{-8}\Kinv$ at 78\K~\cite{Lyon:1977} and $2.59\times10^{-6}\Kinv$ at 298.2\K~\cite{Okada:1984}.
Carbon fiber composites are viable candidates that could satisfy these requirements and have already been used in many high-energy particle physics experiments: At room temperatures, commercially available sheets of 0.5\mm thickness used at the LHC are reported to have thermal conductivity values typically between 400 and $500\WoKm$ and as high as $1100\WoKm$, a Young's modulus of about $400\GPa$, and $\alpha<10^{-6}\Kinv$ along the fiber direction~\cite{CMS:1997tlf}. The exact temperature dependence of the properties of these composite materials are to be measured in the future, and it would be important to use materials that start with the highest thermal conductivity values at room temperature to remain as close to the performance of copper as possible.
For comparison, the thermal conductivity of copper ranges from $\sim400\WoKm$ at room temperature~\cite{CRC:2023} to around $\sim140\WoKm$ at 1\K (for a residual resistance ratio $RRR=30$, with higher $RRR$ resulting in thermal conductivity values much closer to that at room temperature)~\cite{Weisend:1998}. As for durability considerations, the curing process of carbon fiber composites could be adapted for improved microcracking resistance and stability after multiple cryogenic cycles~\cite{Polis:2006}.

The detector setup would also need an external support structure that allows mounting the sensor modules and distributes the cooling power from the copper flange.
We would expect this support structure to have decent thermal conductivity, and its thickness, relative to its radiation length, should be thin enough to avoid significant multiple scattering.
Carbon foam-fiber hybrids and pyrolytic graphite have been considered to hold a large number of pixel detector modules at the LHC and have excellent Young's modulus, high thermal conductivity, and very low $\alpha$ at room temperature~\cite{ATLAS:2017svb,CMS:2017lum}. While the strength and thermal expansion properties are maintained at deep-cryogenic temperatures, their thermal conductivity decreases drastically below $50\K$~\cite{Hauser:1996,Nakamura:2017,Wang:2023}. Inspired by the solution adopted in the ASTRO-H to mitigate excess heat from exposure to the Sun~\cite{Kazuhiro:2018}, we would propose thin, shallow strips of copper with a depth of $\sim1\mm$ to be embedded onto the surface of the support material facing the detector module. The strips could relay the dissipated power from the module to the copper flange to which the support and the ends of the strips would be connected. Using thin strips limits the thermal expansion of the copper area and the stress induced per pixel, and the shallowness of the strips also avoids excessive material and the potential for significant muon multiple scattering.

The assembly of the detector setup is a complex process during which misalignments on the order of the proposed pixel sizes could easily occur. if the top and bottom layers are attached to the same flange, the easiest solution would be to assemble all support structures such that they are interconnected to each other, and install a single large structure under the flange. In this way, laser tests can be performed to measure the positions of individual pixels on the top and bottom layers, and inter-layer misalignments could enter the error correction procedure as calibration constants. Further misalignments that could occur after final commissioning and cooling could be measured in a data-driven way by accumulating enough muon hits to determine additional calibration constants at the level of layers. The same data-driven procedure could be used to determine misalignments between the detector system and the qubit chip. Because a single, large support structure cannot be used when installing the detector layers under two separate flanges, one would most likely need to accumulate more muon interactions to be able to perform alignments at the pixel level in these configurations.

\section{Discussion}
\label{sec:Discussion}

As the physical qubit density increases, the contribution of high-energy particles to errors in quantum computations will become more relevant. Depending on the particular details of the physical setup of the quantum system, muons from cosmic sources may constitute a significant fraction of these contributions.
We have envisioned in this paper the possibility to tag the qubits that are hit by these muons using a two-layer silicon pixel detector and to provide this tagging information as input to a future QEC algorithm that may be able to veto the affected qubits for the duration of their recovery.

The development and implementation of phonon and quasiparticle traps, or membrane structures are necessary to localize upsets over multiple qubits; without such localization, a large portion of the qubit chip is practically disabled, and determining the hit position would not be very useful.
Stricter localization of the upsets is expected to set tighter resolution requirements for hit position predictions over the qubit chip. We find that for resolution requirements within the size of a single qubit, the best muon detection efficiencies are achieved when both detector layers are placed between the cold plate and the qubit chip in a telescopic arrangement with adapted pixel sizes.
Angular acceptance imposed by the dilution refrigerator geometry makes the cold plate and the still flange the only viable candidates to attach the top detector layer for target efficiencies above 50\%.
Loosening the hit position resolution requirement leaves room for the possibility to maintain high detection efficiencies while placing the detector layers on either side of the qubit chip and freeing the bottom layer from the constraints of the dilution refrigerator.
As for the timing resolution of this detector, a reasonable design target would be $\lesssim100\ns$ based on the current QEC round times of $\sim1\mus$.

The still flange and the cold plate are kept at sub-Kelvin temperatures, so the different detector components need to be designed with these low temperatures in mind. Because cooling power at these temperatures is limited, low power consumption is one of the requirements.
Examples to maintain low power from particle physics and space-based experiments include integrating readout electronics in CMOS sensors, specialized readout and amplification schemes, or the employment of double-sided or orthogonal pairs of single-sided strips, which could reduce the channel count drastically and still achieve the desired pixelation. We have also provided an overview of the existing status of and the foreseen challenges for operating undoped silicon and germanium sensors, sensors with light bulk doping profiles, and CMOS technology at $<1\K$. Each of these choices require further refinement in different technological aspects, and experimental demonstration of ultimately maintaining deep-cryogenic, low-power, and reasonably fast operation with infrequent need for maintenance and recalibration.

Ideas concerning the mechanical design and layout of the detector have also been explored in this conceptual study.
While cryogenic tests need to be performed to assess their ultimate usability in this application, carbon fiber composites for modules and carbon-based support structures used in other particle physics experiments appear to be viable candidates for the components of the full detector. Methods to augment heat transfer from the modules may also need to be considered to match the performance of the copper dilution refrigerator disks at deep-cryogenic temperatures.
Finally, once sensor developments are mature enough to begin building full-size layers, the alignment of the pixels will also need to be considered.

Many of the challenges discussed in this paper share common elements with those faced in quantum computing, and particle physics and space-based experiments.
We find that solutions developed in the context of these applications could be adapted for use in this presented case and vice versa. As quantum processors become more advanced in limiting the spread of quasiparticles and phonons over the qubits, we believe the proposed detector setup would be able to tag a significant fraction of muons hits on the qubit chip and contribute to the eventual development of future sensor-assisted QEC algorithms.

\acknowledgments

This project has been supported with the DOE award DE-SC0011702.
We would like to thank Brian Lester (Google Quantum AI) for the valuable discussions on the current understanding of the effects of high-energy particles passing through quantum processors and ideas for future directions to study. We would also like to thank Claudio Campagnari (UCSB) for the valuable questions they raised on the study of muon hit tagging efficiencies and their feedback on this manuscript, Frank Golf (UNL) for discussions on the CMS tracker and possible ways to organize the research program proposed in this paper, Slava Krutelyov (UCSD) for pointing out valuable references on different approaches to lower QEC rates, and Avi Yagil (UCSD) for their encouragement to pursue this study. We would also like to thank Susanne Kyre (UCSB) for thermal interface materials that could be investigated in a full experimental setup, along with ideas to make a mechanical assembly possible.

\bibliographystyle{JHEP}
\bibliography{main.bib}

\appendix
\section{Numerical results of the simulation study for muon detection efficiencies}
\label{app:sec:MuonEffStudyDetails}

In this appendix, we provide the numerical results for the study of muon detection efficiencies with the two-layer pixel detector setup from Section~\ref{sec:MuonEffStudy}.
As noted in that section, the top layer is assumed to be directly below either the cold plate or the still flange in order to maintain $\gtrsim50\%$ muon detection efficiency due to angular acceptance, and the tested values of the width of square pixels in this layer vary between $150$ and $1000\mum$.
The bottom layer is placed either below but away from the cold plate, or over the ground and below the mixing flange, and the pixel sizes in this layer are either kept the same as those for the top layer or adapted for the vertical position of this layer.

When sampling interactions, we assume that all muons from cosmic rays reaching the ground have an incidence angle $\theta$ distributed according to a $\cos^2\theta$ distribution within $\pm60\degrees$. Their momentum is sampled between 0.5 and $10\GeV$ with a distribution according to the experimental data in Ref.~\cite{PDG:2022pth}. While the $\theta$ distribution is known to depend on the momentum of the muon, such dependency should not change the qualitative conclusions of this study. Multiple scattering effects from the different materials in the experimental setup are also included according to Ref.~\cite{PDG:2022pth}.

A selection of results for design choices with reasonable muon detection efficiencies is provided in Table~\ref{table:detparams-cold-plate-cold-plate} for the scenario in which both detector layers are below the cold plate. Only configurations with a pixel area $\geq(500\mum)^2$ in the top layer are listed to keep the numbers of pixels contained inside the dilution refrigerator reasonably low in power consumption. A similar selection of results for design choices pertaining to the scenario in which the bottom detector layer is below the mixing flange instead is provided in Table~\ref{table:detparams-cold-plate-mixing-flange}, and Tables~\ref{table:detparams-still-flange-cold-plate} and \ref{table:detparams-still-flange-mixing-flange} correspond to scenarios in which the top detector layer is underneath the still flange, and the bottom layer is below the cold plate and the mixing flange, respectively.

As also noted in Section~\ref{sec:MuonEffStudy}, it is important to take into account the accuracy constraints from the spread of quasiparticles and phonons. Therefore, following the simplifying assumption that the qubit chip consists of a densely-packed $12\times12$ grid of qubits, each of the Tables~\ref{table:detparams-cold-plate-cold-plate}--\ref{table:detparams-still-flange-mixing-flange} report efficiencies for position accuracies over a $1\times1$ (at the same qubit hit) and $3\times3$ (within nearest neighbors of the qubit hit) qubit cell, in addition to the total efficiency without an accuracy constraint. Efficiencies from a few other possible design choices, including best-case scenarios featuring $(150\mum)^2$-pixels, are also illustrated in a similar manner in Figs.~\ref{fig:effs-top-below-cold-plate} (with top detector layer underneath the cold plate) and \ref{fig:effs-top-below-still-flange} (with top detector layer underneath the still flange). In these figures, efficiencies within a $5\times5$ (within next-to-nearest neighbors of the qubit hit) cell are also displayed for comparison to other qubit cell sizes and can be seen to be not much different from inclusive efficiency values.

\begin{table}[htbp]
\centering
\ccaption
{Parameters of select detector configurations with both layers below the cold plate}
{
The symbols $A_{\text{pixel}}$ and $N_{\text{pixel}}$ represent the area of the square pixels and the product of the number of pixels along the $x$- and $y$-coordinates, respectively. The quantities given in the $\Delta z_{\text{qubit}}^{\text{layer}}$ column are the vertical distances of the top and bottom detector layers, separated by a slash.
Both detector layers are above the qubit chip, so the $\Delta z_{\text{qubit}}^{\text{layer}}$ values are both positive.
A common hit position resolution in the $x$- ($\delta^{95\%}_x$) and $y$- ($\delta^{95\%}_y$) coordinates is reported at 95\%-quantile as the values of the resolution in each coordinate are not significantly different from each other. The values of efficiencies are quoted for correct identification of each qubit ($1\times1$) or when the hit prediction is within the nearest-neighbors of the true qubit hit ($3\times3$), as well as any prediction on the qubit chip.
}
\renewcommand{\arraystretch}{1.25}
\begin{tabular}{ c c c c c }
\hline
$A_{\text{pixel}}$ & \multirow{2}{*}{$N_{\text{pixels}}$} & $\Delta z_{\text{qubit}}^{\text{layer}}$ & $\delta^{95\%}_x$ or $\delta^{95\%}_y$ & Efficiency \\
(${\mumns}^2$) & & ($\mmns$) & ($\mmns$) & 1x1/3x3/any (\%) \\
\hline
$500^{2}$/$220^{2}$ & $390\times400$/$390\times400$ & 242/92 & 0.23 & 64/77/77 \\
$500^{2}$/$500^{2}$ & $390\times400$/$390\times400$ & 242/92 & 0.42 & 52/76/76 \\
$600^{2}$/$260^{2}$ & $325\times333$/$325\times333$ & 242/92 & 0.28 & 61/76/76 \\
$1000^{2}$/$440^{2}$ & $195\times200$/$195\times200$ & 242/92 & 0.48 & 51/77/77 \\
\hdashline
$500^{2}$/$320^{2}$ & $390\times400$/$390\times400$ & 242/142 & 0.53 & 49/76/76 \\
$500^{2}$/$500^{2}$ & $390\times400$/$390\times400$ & 242/142 & 0.70 & 40/76/76 \\
$600^{2}$/$385^{2}$ & $325\times333$/$325\times333$ & 242/142 & 0.64 & 43/76/76 \\
$600^{2}$/$600^{2}$ & $325\times333$/$325\times333$ & 242/142 & 0.85 & 33/76/76 \\
$1000^{2}$/$645^{2}$ & $195\times200$/$195\times200$ & 242/142 & 1.10 & 26/75/76 \\
[1ex]
\hline
\label{table:detparams-cold-plate-cold-plate}
\end{tabular}
\end{table}

\begin{table}[htbp]
\centering
\ccaption
{Parameters of select detector configurations with layers below the cold plate and mixing flange}
{
The symbols $A_{\text{pixel}}$ and $N_{\text{pixel}}$ represent the area of the square pixels and the product of the number of pixels along the $x$- and $y$-coordinates, respectively. The quantities given in the $\Delta z_{\text{qubit}}^{\text{layer}}$ column are the vertical distances of the top and bottom detector layers, separated by a slash.
In these configurations, the top detector layer is directly underneath the cold plate; the bottom detector layer is below the qubit chip, so the second set of $\Delta z_{\text{qubit}}^{\text{layer}}$ values are listed as negative.
A common hit position resolution in the $x$- ($\delta^{95\%}_x$) and $y$- ($\delta^{95\%}_y$) coordinates is reported at 95\%-quantile as the values of the resolution in each coordinate are not significantly different from each other. The values of efficiencies are quoted for correct identification of each qubit ($1\times1$) or when the hit prediction is within the nearest-neighbors of the true qubit hit ($3\times3$), as well as any prediction on the qubit chip.
While the vacuum enclosure and radiation shielding of the dilution refrigerator would place additional constraint on the distance between the bottom detector layer and the bottom surface of the mixing flange,
we nevertheless show all of the tested scenarios to illustrate how efficiencies for the different qubit cell sizes change as the bottom layer approaches closer to the qubit chip from below.
}
\renewcommand{\arraystretch}{1.25}
\begin{tabular}{ c c c c c }
\hline
$A_{\text{pixel}}$ & \multirow{2}{*}{$N_{\text{pixels}}$} & $\Delta z_{\text{qubit}}^{\text{layer}}$ & $\delta^{95\%}_x$ or $\delta^{95\%}_y$ & Efficiency \\
(${\mumns}^2$) & & ($\mmns$) & ($\mmns$) & 1x1/3x3/any (\%) \\
\hline
$500^{2}$/$500^{2}$ & $390\times400$/$551\times564$ & 242/-308 & 1.05 & 41/74/77 \\
$500^{2}$/$705^{2}$ & $390\times400$/$390\times400$ & 242/-308 & 1.05 & 41/74/77 \\
$600^{2}$/$600^{2}$ & $325\times333$/$460\times470$ & 242/-308 & 1.05 & 41/74/77 \\
$600^{2}$/$850^{2}$ & $325\times333$/$325\times333$ & 242/-308 & 1.06 & 40/74/77 \\
$1000^{2}$/$1000^{2}$ & $195\times200$/$276\times282$ & 242/-308 & 1.08 & 38/74/77 \\
$1000^{2}$/$1415^{2}$ & $195\times200$/$195\times200$ & 242/-308 & 1.10 & 36/74/77 \\
\hdashline
$500^{2}$/$500^{2}$ & $390\times400$/$390\times400$ & 242/-208 & 0.86 & 45/75/77 \\
$600^{2}$/$600^{2}$ & $325\times333$/$325\times333$ & 242/-208 & 0.87 & 45/75/77 \\
$1000^{2}$/$1000^{2}$ & $195\times200$/$195\times200$ & 242/-208 & 0.91 & 41/75/77 \\
\hdashline
$500^{2}$/$270^{2}$ & $390\times400$/$390\times400$ & 242/-108 & 0.56 & 54/76/76 \\
$500^{2}$/$500^{2}$ & $390\times400$/$390\times400$ & 242/-108 & 0.58 & 52/76/77 \\
$600^{2}$/$320^{2}$ & $325\times333$/$325\times333$ & 242/-108 & 0.57 & 54/76/76 \\
$600^{2}$/$600^{2}$ & $325\times333$/$325\times333$ & 242/-108 & 0.59 & 51/76/76 \\
$1000^{2}$/$535^{2}$ & $195\times200$/$195\times200$ & 242/-108 & 0.60 & 51/76/76 \\
$1000^{2}$/$1000^{2}$ & $195\times200$/$195\times200$ & 242/-108 & 0.66 & 46/76/77 \\
\hdashline
$500^{2}$/$160^{2}$ & $390\times400$/$390\times400$ & 242/-58 & 0.34 & 62/77/77 \\
$500^{2}$/$500^{2}$ & $390\times400$/$390\times400$ & 242/-58 & 0.38 & 58/76/76 \\
$600^{2}$/$190^{2}$ & $325\times333$/$325\times333$ & 242/-58 & 0.35 & 62/77/77 \\
$600^{2}$/$600^{2}$ & $325\times333$/$325\times333$ & 242/-58 & 0.40 & 56/76/76 \\
$1000^{2}$/$315^{2}$ & $195\times200$/$195\times200$ & 242/-58 & 0.37 & 60/77/77 \\
$1000^{2}$/$1000^{2}$ & $195\times200$/$195\times200$ & 242/-58 & 0.51 & 48/76/76 \\
[1ex]
\hline
\label{table:detparams-cold-plate-mixing-flange}
\end{tabular}
\end{table}

\begin{table}[htbp]
\centering
\ccaption
{Parameters of select detector configurations with layers below the still flange and cold plate}
{
The symbols $A_{\text{pixel}}$ and $N_{\text{pixel}}$ represent the area of the square pixels and the product of the number of pixels along the $x$- and $y$-coordinates, respectively. The quantities given in the $\Delta z_{\text{qubit}}^{\text{layer}}$ column are the vertical distances of the top and bottom detector layers, separated by a slash.
In these configurations, the top detector layer is directly underneath the still flange, and the bottom detector layer is between the cold plate and the qubit chip.
A common hit position resolution in the $x$- ($\delta^{95\%}_x$) and $y$- ($\delta^{95\%}_y$) coordinates is reported at 95\%-quantile as the values of the resolution in each coordinate are not significantly different from each other. The values of efficiencies are quoted for correct identification of each qubit ($1\times1$) or when the hit prediction is within the nearest-neighbors of the true qubit hit ($3\times3$), as well as any prediction on the qubit chip.
}
\renewcommand{\arraystretch}{1.25}
\begin{tabular}{ c c c c c }
\hline
$A_{\text{pixel}}$ & \multirow{2}{*}{$N_{\text{pixels}}$} & $\Delta z_{\text{qubit}}^{\text{layer}}$ & $\delta^{95\%}_x$ or $\delta^{95\%}_y$ & Efficiency \\
(${\mumns}^2$) & & ($\mmns$) & ($\mmns$) & 1x1/3x3/any (\%) \\
\hline
$500^{2}$/$140^{2}$ & $390\times400$/$390\times400$ & 419/92 & 0.38 & 41/52/52 \\
$500^{2}$/$500^{2}$ & $390\times400$/$390\times400$ & 419/92 & 0.47 & 35/52/52 \\
$600^{2}$/$165^{2}$ & $325\times333$/$325\times333$ & 419/92 & 0.39 & 40/52/52 \\
$1000^{2}$/$280^{2}$ & $195\times200$/$195\times200$ & 419/92 & 0.42 & 38/52/52 \\
\hdashline
$500^{2}$/$200^{2}$ & $390\times400$/$390\times400$ & 419/142 & 0.69 & 33/52/52 \\
$600^{2}$/$240^{2}$ & $325\times333$/$325\times333$ & 419/142 & 0.70 & 33/52/52 \\
$1000^{2}$/$400^{2}$ & $195\times200$/$195\times200$ & 419/142 & 0.76 & 30/52/52 \\
[1ex]
\hline
\label{table:detparams-still-flange-cold-plate}
\end{tabular}
\end{table}

\begin{table}[htbp]
\centering
\ccaption
{Parameters of select detector configurations with layers below the still and mixing flanges}
{
The symbols $A_{\text{pixel}}$ and $N_{\text{pixel}}$ represent the area of the square pixels and the product of the number of pixels along the $x$- and $y$-coordinates, respectively. The quantities given in the $\Delta z_{\text{qubit}}^{\text{layer}}$ column are the vertical distances of the top and bottom detector layers, separated by a slash.
In these configurations, the top detector layer is directly underneath the still flange; the bottom detector layer is below the qubit chip, so the second set of $\Delta z_{\text{qubit}}^{\text{layer}}$ values are listed as negative.
A common hit position resolution in the $x$- ($\delta^{95\%}_x$) and $y$- ($\delta^{95\%}_y$) coordinates is reported at 95\%-quantile as the values of the resolution in each coordinate are not significantly different from each other. The values of efficiencies are quoted for correct identification of each qubit ($1\times1$) or when the hit prediction is within the nearest-neighbors of the true qubit hit ($3\times3$), as well as any prediction on the qubit chip.
While the vacuum enclosure and radiation shielding of the dilution refrigerator would place additional constraint on the distance between the bottom detector layer and the bottom surface of the mixing flange,
we nevertheless show all of the tested scenarios to illustrate how efficiencies for the different qubit cell sizes change as the bottom layer approaches closer to the qubit chip from below.
}
\renewcommand{\arraystretch}{1.25}
\begin{tabular}{ c c c c c }
\hline
$A_{\text{pixel}}$ & \multirow{2}{*}{$N_{\text{pixels}}$} & $\Delta z_{\text{qubit}}^{\text{layer}}$ & $\delta^{95\%}_x$ or $\delta^{95\%}_y$ & Efficiency \\
(${\mumns}^2$) & & ($\mmns$) & ($\mmns$) & 1x1/3x3/any (\%) \\
\hline
$500^{2}$/$500^{2}$ & $390\times400$/$390\times400$ & 419/-308 & 1.44 & 23/48/52 \\
$600^{2}$/$600^{2}$ & $325\times333$/$325\times333$ & 419/-308 & 1.44 & 23/48/52 \\
$1000^{2}$/$840^{2}$ & $195\times200$/$195\times200$ & 419/-308 & 1.46 & 22/48/52 \\
\hdashline
$500^{2}$/$500^{2}$ & $390\times400$/$390\times400$ & 419/-208 & 1.13 & 27/50/52 \\
$600^{2}$/$600^{2}$ & $325\times333$/$325\times333$ & 419/-208 & 1.14 & 26/50/52 \\
$1000^{2}$/$590^{2}$ & $195\times200$/$195\times200$ & 419/-208 & 1.14 & 26/50/52 \\
\hdashline
$500^{2}$/$170^{2}$ & $390\times400$/$390\times400$ & 419/-108 & 0.68 & 35/51/52 \\
$500^{2}$/$500^{2}$ & $390\times400$/$390\times400$ & 419/-108 & 0.70 & 33/51/52 \\
$600^{2}$/$200^{2}$ & $325\times333$/$325\times333$ & 419/-108 & 0.68 & 34/51/52 \\
$600^{2}$/$600^{2}$ & $325\times333$/$325\times333$ & 419/-108 & 0.71 & 32/51/52 \\
$1000^{2}$/$335^{2}$ & $195\times200$/$195\times200$ & 419/-108 & 0.69 & 34/51/52 \\
\hdashline
$500^{2}$/$500^{2}$ & $390\times400$/$390\times400$ & 419/-58 & 0.44 & 38/52/52 \\
$600^{2}$/$125^{2}$ & $325\times333$/$325\times333$ & 419/-58 & 0.39 & 41/52/52 \\
$600^{2}$/$600^{2}$ & $325\times333$/$325\times333$ & 419/-58 & 0.46 & 36/52/52 \\
$1000^{2}$/$210^{2}$ & $195\times200$/$195\times200$ & 419/-58 & 0.40 & 40/52/52 \\
$1000^{2}$/$1000^{2}$ & $195\times200$/$195\times200$ & 419/-58 & 0.57 & 31/52/52 \\
[1ex]
\hline
\label{table:detparams-still-flange-mixing-flange}
\end{tabular}
\end{table}

\begin{figure*}[htb]
\centering
\includegraphics[width=0.9\textwidth]{c\_top\_below\_Cold\_plate\_bottom\_below\_Cold\_plate\_-150mm\_accuracyCellSize.pdf} \\
\includegraphics[width=0.9\textwidth]{c\_top\_below\_Cold\_plate\_bottom\_below\_Mixing\_flange\_-300mm\_accuracyCellSize.pdf} \\
\ccaption
{Efficiencies for select detector configurations with the top layer underneath the cold plate}
{
The efficiencies are shown for accurate hit identification within $1\times1$, $3\times3$, or $5\times5$ qubit grid cell sizes as well as for any level of accuracy with a valid hit prediction over the qubit chip. The bottom detector layer is $150\mm$ below the cold plate in configurations displayed on the top panel, and $300\mm$ below the mixing flange in those displayed on the bottom panel. The arm length between the detector layers is long enough in the latter scenario that multiple scattering from the mixing flange reduces muon detection efficiencies for $3\times3$ and $5\times5$ qubit cells visibly. The different colors correspond to different pixel sizes in the top detector layer. Configurations corresponding to the scenario in which the bottom layer consists of pixels of the same size are shown with solid histograms, and those with pixel sizes adapted to the position of the bottom layer are shown in dashed histograms. The sizes and the numbers of pixels in each configuration are displayed on the legend at the right.
}
\label{fig:effs-top-below-cold-plate}
\end{figure*}

\begin{figure*}[htb]
\centering
\includegraphics[width=0.9\textwidth]{c\_top\_below\_Still\_flange\_bottom\_below\_Cold\_plate\_-150mm\_accuracyCellSize.pdf} \\
\includegraphics[width=0.9\textwidth]{c\_top\_below\_Still\_flange\_bottom\_below\_Mixing\_flange\_-300mm\_accuracyCellSize.pdf} \\
\ccaption
{Efficiencies for select detector configurations with the top layer underneath the still flange}
{
The efficiencies are shown for accurate hit identification within $1\times1$, $3\times3$, or $5\times5$ qubit grid cell sizes as well as for any level of accuracy with a valid hit prediction over the qubit chip. The bottom detector layer is $150\mm$ below the cold plate in configurations displayed on the top panel, and $300\mm$ below the mixing flange in those displayed on the bottom panel. The arm length between the detector layers is long enough in the latter scenario that multiple scattering from the cold plate and mixing flange compounds to reduce muon detection efficiencies for $3\times3$ and $5\times5$ qubit cells visibly. The different colors correspond to different pixel sizes in the top detector layer. Configurations corresponding to the scenario in which the bottom layer consists of pixels of the same size are shown with solid histograms, and those with pixel sizes adapted to the position of the bottom layer are shown in dashed histograms. The sizes and the numbers of pixels in each configuration are displayed on the legend at the right.
}
\label{fig:effs-top-below-still-flange}
\end{figure*}

\end{document}